\documentclass[conference,compsoc]{IEEEtran}
\IEEEoverridecommandlockouts
\usepackage{cite}
\usepackage{amsmath,amssymb,amsfonts}
\usepackage{algorithmic}
\usepackage{graphicx}
\usepackage{textcomp}
\usepackage{xcolor}
\usepackage{booktabs} 
\usepackage{url}
\usepackage{tabularx}
\usepackage{enumitem}
\usepackage[breakable]{tcolorbox}
\tcbuselibrary{breakable}
\usepackage{enumitem}
\usepackage{hyperref}
\newtcolorbox[auto counter, number within=subsection]{promptbox}[2][]{
    floatplacement=t,   
    float,
    breakable,
    colback=white,
    colframe=black,
    fonttitle=\bfseries,
    title={Prompt~\thetcbcounter: #2},
    boxrule=0.8mm,
    before skip=10pt,
    after skip=10pt,
    #1
}

\newtcolorbox[auto counter, number within=subsection]{commentbox}[2][]{
    floatplacement=t,   
    float,
    breakable,
    colback=white,
    colframe=black,
    fonttitle=\bfseries,
    title={Comment~\thetcbcounter: #2},
    boxrule=0.8mm,
    before skip=10pt,
    after skip=10pt,
    #1
}

\newtcolorbox[auto counter, number within=subsection]{codebox}[1][]{
    floatplacement=t,   
    float,
    breakable,
    colback=white,
    colframe=black,
    fonttitle=\bfseries,
    title={Code~\thetcbcounter: #1},
    boxrule=0.8mm,
    before skip=10pt,
    after skip=10pt,
}

\usepackage{listings}
\usepackage{xcolor}

\lstdefinelanguage{asm}{
  morekeywords=[1]{mov,add,sub,jmp,cmp,je,jne,int,ret,lea,push,pop,call,and,or,shl,shr,xchg,neg,xor},
  morekeywords=[2]{eax,ebx,ecx,edx,esi,edi,esp,ebp,ax,bx,cx,dx,al,ah,bl,bh,cl,ch,dl,dh,rax,rbx,rcx,rdx},
  sensitive=true,
  morecomment=[l]{;},
  morestring=[b]"
}

\def\BibTeX{{\rm B\kern-.05em{\sc i\kern-.025em b}\kern-.08em
    T\kern-.1667em\lower.7ex\hbox{E}\kern-.125emX}}

\begin{document}
\title{REx86: A Local Large Language Model for Assisting in x86 Assembly Reverse Engineering}


\author{
\IEEEauthorblockN{Darrin Lea, James Ghawaly, Golden Richard III, Aisha Ali-Gombe, and Andrew Case}
\IEEEauthorblockA{
Division of Computer Science and Engineering, Louisiana State University, Baton Rouge, USA\\
\texttt{\{dlea1, jghawaly, goldenrichard1, aaligombe, andrewcase\}@lsu.edu}
}
}

\maketitle
\IEEEpubid{\begin{minipage}{\linewidth}
\centering\footnotesize
© 2025 IEEE. Personal use of this material is permitted. Permission from IEEE must be obtained for all other uses.
\end{minipage}}

\begin{abstract}


Reverse engineering (RE) of x86 binaries is indispensable for malware and firmware analysis, but remains slow due to stripped metadata and adversarial obfuscation. Large Language Models (LLMs) offer potential for improving RE efficiency through automated comprehension and commenting, but cloud-hosted, closed-weight models pose privacy and security risks and cannot be used in closed-network facilities. We evaluate parameter-efficient fine-tuned local LLMs for assisting with x86 RE tasks in these settings. Eight open-weight models across the CodeLlama, Qwen2.5-Coder, and CodeGemma series are fine-tuned on a custom curated dataset of 5,981 x86 assembly examples. We evaluate them quantitatively and identify the fine-tuned Qwen2.5-Coder-7B as the top performer, which we name REx86.

REx86 reduces test-set cross-entropy loss by 64.2\% and improves semantic cosine similarity against ground truth by 20.3\% over its base model. In a limited user case study (n=43), REx86 significantly enhanced line-level code understanding (p = 0.031) and increased the correct-solve rate from 31\% to 53\% (p = 0.189), though the latter did not reach statistical significance. Qualitative analysis shows more accurate, concise comments with fewer hallucinations.

REx86 delivers state-of-the-art assistance in x86 RE among local, open-weight LLMs. Our findings demonstrate the value of domain-specific fine-tuning, and highlight the need for more commented disassembly data to further enhance LLM performance in RE. REx86, its dataset, and LoRA adapters are publicly available at https://github.com/dlea8/REx86 and https://zenodo.org/records/15420461.



\end{abstract}


\section{Introduction}\label{sec:intro}


Reverse engineering (RE), the process of discovering and understanding the functionality of a piece of software, is frequently used by cybersecurity practitioners as a tactic to uncover the structure of malware or identify vulnerabilities in software and firmware \cite{idabook}\cite{chikofsky2002reverse}\cite{muller2000reverse}. However, RE is a complex, tedious, and time-consuming process for several reasons. For one, compilation removes descriptors and designations, such as user-defined datatypes, variable names, and comments, which are taken for granted in most development environments \cite{idabook}. Furthermore, compiler optimizations increase code efficiency at the cost of readability. Working with malware adds another level of complexity, as malware authors often intentionally obfuscate code to hinder RE efforts \cite{mohseni2024can} and use esoteric programming techniques to achieve their goals. 

Reverse engineering tools, such as IDA Pro \cite{ida} and Ghidra \cite{ghidra}, offer disassemblers, decompilers, and various other features for assisting with RE. Still, none can resurrect the code documentation lost through compilation. 

Large Language Models (LLMs) offer a potential solution to the lack of documentation during RE because of their aptitude for generating natural language \cite{zhou2023lima}. While LLMs have proven effective in other fields of cybersecurity, such as malware classification \cite{walton2025exploring}\cite{qian2025lamd}\cite{sanchez2024transfer}, IoC identification \cite{sharma2025forensicllm}, TTP analysis \cite{tseng2024using}\cite{williamson2024malware}, and digital forensics \cite{sharma2025forensicllm}, fewer research efforts have explored the effectiveness of using local LLMs to document malware samples during an RE effort. One reason for this gap is that pre-trained LLMs struggle to comprehend low-level code, particularly on a broad scale. LLMs can effectively describe the functionality of specific low-level instructions but fail to comprehend the purpose of that instruction in the wider context of the program. This points to a lack of exposure to low-level code in the training process of LLMs, even for code-oriented models. Thus, there is a need to extend the capabilities of current LLMs.


\subsection{Research Questions and Contributions}

In this paper, we investigate the application of custom, local, and open-weight LLMs to x86 RE. Specifically, we seek to address the following research questions:

\begin{enumerate}[leftmargin=*,labelindent=0.6cm]
\itemsep0em
\item[\textbf{RQ1}] Which local LLM architectures and parameter sizes are most effective at generating accurate comments for disassembly output after fine-tuning on x86-specific datasets?
\item[\textbf{RQ2}] To what extent can fine-tuned, local, open-weight LLMs improve the understanding of x86 assembly code for reverse engineering tasks?
\end{enumerate}

In addressing these research questions, our work makes the following contributions to the community:

\begin{itemize}
    \item \textbf{REx86 LLM Weights} - All fine-tuned models are publicly available at \url{https://zenodo.org/records/15420461} as LoRA adapters, including the highest-performing model, REx86. Our models are intentionally designed for efficient operation on consumer-grade GPUs. Being efficient, local, and open-weight, we encourage the community to use and improve them. 
    \item \textbf{REx86 Assembly Dataset} - Section \ref{dataset_curation} outlines the curation of an x86 assembly dataset that is used to fine-tune our models. This dataset, publicly available at \url{https://github.com/dlea8/REx86}, could be a useful resource for other reverse engineering and malware analysis research. 
    \item \textbf{Quantitative, Qualitative, and Human Evaluation of REx86} - We analyze eight fine-tuned models, comparing their effectiveness quantitatively in various RE-related tasks against their baseline counterparts. We also performed a qualitative evaluation of the best model, REx86, on various code samples. Finally, we performed a human evaluation study to evaluate its usefulness in a static malware analysis task. 
\end{itemize}

\subsection{Intended Deployment Context}
Many RE programs operate inside connectivity-restricted enclaves where third-party APIs are unavailable or prohibited (e.g., SCIFs governed by ICD/ICS-705\cite{NCSC_SCIF_TechSpecs_v1_5_2020}; DoD enclaves requiring DISA SRG IL5/IL6 \cite{DISA_Cloud_Computing_SRG_V1R4_2022}; OT/ICS incident response zones aligned with NIST SP 800-82\cite{NIST_SP_800_82r3_2023}; organizations handling CUI under NIST SP 800-171\cite{NIST_SP_800_171r3_2024}). In these settings, analysts need tools that run entirely on-prem, leave no data egress, and are inspectable. REx86 is designed for this reality: it is local, open-weight, and runs at full precision on a single high-end consumer GPU, making it deployable in enclaves where cloud LLMs are not an option. Our goal is not to outperform cloud-based frontier models in aggregate, but to deliver a capable, local baseline that materially assists analysts where closed-source API models cannot be used. 




\subsection{Background}

\subsubsection{Reverse Engineering}

At a high level, RE refers to discovering the functionality of a piece of software, malware, or firmware \cite{idabook}\cite{chikofsky2002reverse}. This often involves extracting human-readable information from a compiled binary or executable file \cite{idabook}. The compilation process starts when source code written in a human-readable programming language is compiled into an assembly language, such as the x86 instruction set. During compilation, the code is optimized for performance at the cost of readability. Variable names, comments, and user-defined datatypes are considered extraneous to the compiler and are removed \cite{idabook}. Finally, assemblers and linkers convert the assembly code into machine code, creating the executable file \cite{idabook}. 

Reverse engineering unwinds this compilation process using disassemblers and decompilers. Given an executable, the machine code is converted back to assembly code using a disassembler \cite{idabook}. Many RE tools like Ghidra and IDA Pro offer a decompiler that converts assembly code into a C-based pseudocode \cite{ida} \cite{ghidra}. However, complete source code recovery is impossible due to the loss of documentation, user-defined types, and variable names during compilation \cite{idabook}.

\subsubsection{Fine-tuning}

LLMs are trained on a large corpus of knowledge and perform well on a wide variety of tasks \cite{hu2023survey}\cite{zhou2023lima}. However, oftentimes there is a need to extend the functionality of an LLM to encompass domain-specific knowledge or capabilities \cite{hu2023survey}\cite{dos2024domain}\cite{aghajanyan2020intrinsic}. One of the most popular ways to expand the functionality of an LLM is by fine-tuning \cite{dos2024domain}. Fine-tuning embeds new information into the model itself by updating the weights through additional training on a domain-specific dataset \cite{ovadia2023FTorRAG}\cite{soudani2024fTvRAG}. 

\subsubsection{Low-Rank Adaptation}

Fine-tuning can be resource-intensive and time-consuming, especially when training larger models. Low-rank adaptation (LoRA) is a Parameter Efficient Fine-Tuning (PEFT) solution that reduces the number of trainable parameters during fine-tuning, greatly decreasing the computational resources required \cite{hu2022lora}. LoRA interprets a fine-tuned weight matrix in a model as a combination of the base model's weight matrix and a fine-tuned low-rank update matrix. For any given fine-tuned weight matrix in the model, the equation\[
W' = W + \Delta W
\] represents the fine-tuned weight matrix, $W'\in \mathcal{R}^{M \times N}$, comprising an additive combination of the base model weight matrix, $W\in \mathcal{R}^{M \times N}$, and $\Delta W\in \mathcal{R}^{M \times N}$, the weight update matrix optimized through fine-tuning. $\Delta W$ is decomposed into two matrices, $A\in \mathcal{R}^{M \times r}$ and $B\in \mathcal{R}^{r \times N}$, where $r$ is the rank of the update matrix. 
The rank is typically chosen as a power of two between 8 and 64. For most fine-tuning applications where the new task is similar to the base model's capabilities, increasing $r$ beyond roughly 32 does not yield significant improvement gains. A second hyperparameter, $\alpha$, is used as a scaling factor to tune the impact of the LoRA weight matrix. The standard heuristic is to set $\alpha=r$ or $\alpha=2r$. 

Thus, the full LoRA formula to represent a fine-tuned model is given below.\[
W' = W + \frac{\alpha}{r}(AB)
\]

\subsubsection{Quantization}

Quantization is a method of reducing model size by decreasing the precision of the numerical weights of the model \cite{frantar2022optq}. Precision is usually reduced from 16-bit to 8-bit or 4-bit. While quantization can result in 1-3\% performance reduction, the benefit of being able to load a larger model often outweighs this small reduction in accuracy \cite{frantar2022optq}\cite{jin2024comprehensive}\cite{dettmers2023case}. It has been shown that 4-bit precision is typically optimal for balancing loss of accuracy and model memory efficiency \cite{dettmers2023case}. 

Unsloth, the fine-tuning framework used in this paper, utilizes 4-bit quantization based on the bitsandbytes scheme \cite{unsloth}. This scheme uses post-training quantization with the 4-bit NormalFloat (nf4) datatype introduced by QLoRA \cite{roy2023understanding}\cite{dettmers2023qlora}. The 14B and 32B parameter models were fine-tuned using this 4-bit quantization to enable them to fit on consumer hardware.




\section{Prior Work}\label{sec:priorwork}

The use of LLMs to expedite RE has become increasingly popular as a research topic in recent years. Many have utilized the wide range of competence of LLMs for high-level malware analysis. For example, Williamson uses Gemini Pro to identify indicators of compromise (IoC) in software and records promising results \cite{williamson2024malware}. Another paper employs LLMs to generate Cyber Threat Intelligence (CTI) graphs \cite{tseng2024using}. Additionally, Fayyazi et al. utilize LLMs to identify tactics, techniques, and procedures (TTP) used in malware \cite{fayyazi2024advancing}.


Additionally, several works focus on lower-level RE, optimizing output of IDA's and Ghidra's decompilers. Some approaches aim to summarize source code \cite{al2023extending}\cite{ahmad2021transformer}. Walton et al. adopt a tiered approach to source code summarization of Android Malware by annotating at the class level and then aggregating class comments into full module summarizations \cite{walton2025exploring}. LAMD uses a similar tiered approach for code summarization, but aims to extract only security-critical sections of code \cite{qian2025lamd}. DeGPT refines the output of the decompiler by assigning useful variable names, generating comments, and deobfuscating code structure \cite{hu2024degpt}. LLM4Decompile explores direct decompilation of binary code using LLMs \cite{tan2024llm4decompile}. The authors discover the LLM decompilation outperforms IDA and Ghidra, but is inferior to post-decompilation refinements \cite{tan2024llm4decompile}. 


The works described above focus on optimizing decompilation output, but additional research efforts attempt to boost assembly code comprehensibility. While we will mainly focus on LLM approaches in this paper, it should be noted that researchers have also presented approaches that automate reverse engineering through other machine learning techniques. For example, Anderson et al. use machine learning to classify and label assembly subroutines based on instructions, API calls, and other nearby subroutines \cite{anderson2014automateRE}.


 Another collection of work focuses on using vector embeddings to foster assembly code comprehension. For example, Asm2Vec attempts to capture the semantic meanings of assembly instructions in vector embeddings \cite{ding2019asm2vec}. A similar approach is taken to create BinBert \cite{artuso2024binbert} and PalmTree \cite{li2021palmtree}.


LLMs have also proven their ability to generate assembly code. In their work, Mohseni et al. find success with generating obfuscated assembly code with a variety of LLM series \cite{mohseni2024can}. Additionally, while they don't use LLMs, Liguori et al. show that machine learning is effective in generating assembly shellcode snippets \cite{ShellcodeIA32}. 


Finally, ASMA-Tune is a framework that separates the sematic assembly instructions comprehension and natural language generation into separate modules \cite{wang2025asma}. In this paper, the authors use an encoder module for capturing semantic meaning of assembly instructions and a decoder LLM for generating natural language about assembly architectures. Their model is successful in boosting assembly code comprehension \cite{wang2025asma}. However, they rely heavily on synthetic data to train their model. Additionally, they only fine-tune one model, whereas we survey several models across multiple LLM series.

\section{Methodology}\label{sec:methodology}

\subsection{Dataset Curation}
\label{dataset_curation}

\begin{figure*}
    \centering
    \includegraphics[width=\textwidth]{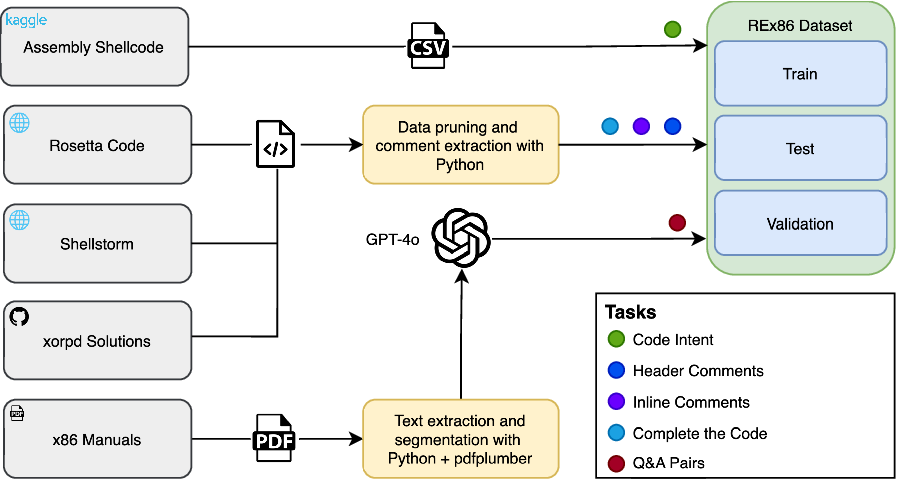}
    \caption{The process of curating the x86 fine-tuning dataset}
    \label{fig:dataset-curation}
\end{figure*}

The fine-tuning dataset focuses on five tasks that facilitate a deeper understanding of x86 assembly code. The x86 code samples are acquired from four online repositories and question-answer pairs are extracted from several x86 assembly manuals and textbooks. The dataset is formatted in the common Alpaca style, containing an instruction, input, and output for each entry. During training, the model is provided the instruction and input, and its response is checked against the provided output for that entry. 

Note that our dataset only contains entries related to the x86 architecture. While there are numerous architectures that power the countless devices in circulation today, this paper focuses on x86 for several reasons. First, x86 is the underlying architecture for Windows and many Linux devices, making it the primary architecture for most laptops, desktops, and servers. Thus, much of the software and malware written today is compiled into x86 assembly to run on these machines. Additionally, numerous repositories of x86 code are available online due to the prevalence of the architecture. Finally, while we intend to create a generalized model in future work, doing so would likely prove impossible without first improving comprehension of a single architecture. Thus, we select x86 as the architecture to assess the research questions \textbf{RQ1} and \textbf{RQ2} introduced in Section \ref{sec:intro}.

\subsubsection{Tasks}

The fine-tuning dataset consists of five tasks: Code Intent, Complete the Code, Inline Comments, Header Comment, and Q\&A. Each task type provides a unique angle for understanding x86 assembly. 

The Code Intent samples challenge the model to reason about the purpose of the code. For Code Intent entries, the model is provided with the code snippet, and its response is checked against the intent for that entry. 

The Complete the Code entries are intended to enhance the model's ability to write x86 assembly code. For these samples, the model is provided the code sample with 25\% of lines masked and is instructed to complete the code by filling in the appropriate instruction for each masked line. The model's response is compared with the complete code sample. 

Header Comment samples facilitate better code summarization. The model is provided a code sample as input and is instructed to generate a header comment for the sample, which is compared with the header comment from the dataset entry.

The Inline Comments samples challenge the model to reason about an instruction's role in the context of the program rather than its surface-level functionality. The model is provided a full code sample as input and is instructed to output a JSON object of line numbers and the corresponding comment for each line number.

The Q\&A entries train the model's ability to communicate effectively about the x86 architecture itself. the model is given a question and is instructed to provide an answer. Its answer is checked against the dataset entry. 

The prompts for each task type are provided below. 

\begin{itemize}
    \item \textbf{Code Intent} — 
    \begin{quote}
        \texttt{Describe the intent of the given snippet of assembly code.}
    \end{quote}

    \item \textbf{Complete the Code} — 
    \begin{quote}
        \texttt{An x86 Assembly code snippet has been partially masked. Complete the code by filling in the lines labeled '\# <MASKED>'.}
    \end{quote}

    \item \textbf{Inline Comments} — 
    \begin{quote}
        \texttt{Comment the x86 assembly code snippet by generating a structured JSON object where the keys are the integer line numbers (starting at 1) and the values are the string comments. For example: \{1: 'comment for line 1', 2: 'comment for line 2', 3: 'comment for line 3'\}.}
    \end{quote}

    \item \textbf{Header Comment} — 
    \begin{quote}
        \texttt{Write a header comment for this x86 assembly code snippet.}
    \end{quote}

    \item \textbf{Q\&A} — 
    \begin{quote}
        \texttt{Answer the following question about the x86 Assembly Language.}
    \end{quote}
\end{itemize}

\subsubsection{Data Sources}

Due to the lack of a comprehensive x86 assembly dataset containing commented code, we select four sources from which to aggregate x86 code samples to form a fine-tuning dataset: Assembly Shellcode Dataset \cite{kaggleAssemblyShellcode}, Rosetta Code \cite{rosettacode}, Shell-Storm \cite{shellstorm}, and xorpd Solutions \cite{xorpdsolutions}. 


The Assembly Shellcode Dataset is a pre-curated dataset of roughly 3000 x86 assembly code snippets and the corresponding intent of each snippet. All Code Intent samples in the fine-tuning dataset are taken from this source. The Assembly Shellcode dataset provides a solid starting point, but ultimately lacks the volume and diversity needed to stand on its own. Thus, we supplement it with several additional sources of x86 code.

Rosetta Code is a repository of solutions to coding problems written in various programming languages, including x86 assembly. Approximately one hundred x86 samples, including header-commented, inline-commented, and uncommented code, are downloaded from Rosetta Code. 

Another three hundred samples are extracted from Shell-Storm, an online aggregation of x86 shellcode examples. These samples also consist of header-commented code, inline-commented code, and uncommented code. 

Additionally, roughly one hundred sixty commented samples are downloaded from the xorpd-solutions Github repository, a collection of answers to the x86 samples in the \textit{xchg rax,rax} book \cite{xorpdschg}. This repository contains both header comments and inline comments for every sample. Thus, each sample can be utilized as a header-commented sample by removing the inline comments or as an inline-commented sample by removing the header comments. 

Finally, to supplement the code samples gathered from the previous four sources, question-answer pairs are extracted from x86 manuals and textbooks using OpenAI's GPT-4o model. The sources from which Q\&A pairs are extracted from are listed below. 

\begin{itemize}
    \item PC Assembly Language by Paul A. Carter \cite{carter2006pcassembly}
    \item Oracle's x86 Assembly Language Reference Manual \cite{oracle1995x86}
    \item SunSoft's x86 Assembly Language Reference Manual \cite{oracle2005solaris}
    \item University of Virginia's x86 Assembly Guide \cite{evansx86virginia}
    \item Assembly Language for x86 Processors Chapter 5: x86 Assembly Language \cite{irvine2010procedures}
\end{itemize}

Each manual is segmented into sections, and GPT-4o is instructed to generate up to five question-answer pairs pertaining to each section. The prompt provided to GPT-4o is shown in Prompt~\ref{box:qa_extract}.

\begin{promptbox}[label={box:qa_extract}]{Q\&A Extraction Prompt}\label{prompt:extraction}
The above text is an excerpt from a section of an x86 Assembly Language manual. Your job is to extract a set of up to 5 question-and-answer pairs. 
Five questions is an upper limit, but it is not necessary to generate five questions if the content does not support it.
The questions should be answerable without knowing what manual the questions are based on.
    
\textbf{Important Restrictions}:
\begin{itemize}
    \item The questions should strictly focus on the \textbf{technical content} in the section.
    \item Do \textbf{not} generate questions about the manual itself (e.g., its purpose, target audience, structure, or authorship).
    \item Do \textbf{not} reference the manual in any form (e.g., avoid phrases like ``this guide'', ``this manual'',``this section'', or ``this document'').
    \item The questions must \textbf{only} ask about the actual concepts, instructions, or principles found in the provided section.
\end{itemize}
    
\textbf{Guidelines}:
\begin{enumerate}
    \item \textbf{Relevance}:  
    \begin{itemize}
        \item Ensure questions are strictly about the \textbf{technical content} in the section. 
        \item Do \textbf{not} generate questions regarding the \textbf{manual itself} (e.g., its focus, purpose, or intended audience). 
        \item Example of what \textbf{NOT} to ask: ``What is the purpose of this manual?'' 
        \item Example of a \textbf{good question}: ``What are the different types of MOV instructions in x86 assembly?'' 
    \end{itemize}
    \item \textbf{Language Use}:  
    \begin{itemize}
        \item Formulate \textbf{clear and concise} questions.  
        \item Answers must use language from the section as much as possible to maintain accuracy and context.
    \end{itemize}
    \item \textbf{Answer Quality}
    \begin{itemize}
        \item Answers must be \textbf{detailed} and provide \textbf{comprehensive} explanations.
        \item The length of answers should vary based on the complexity of the question. 
    \end{itemize}
\end{enumerate}

\end{promptbox}

With samples aggregated from these five sources, the fine-tuning dataset contains a total 5981 entries spanning the five tasks and is available at \url{https://github.com/dlea8/REx86}. Table \ref{tab:dataset_task_type_breakdown} shows a breakdown of the dataset by number of entries for each task. For fine-tuning, 70\% of the dataset is allocated for training, 10\% for validation, and 20\% for testing. Each segment contains a proportional split of each of the five task types. 

\begin{table}[h]
    \centering
    \caption{Number of dataset entries broken down by task type}
    \begin{tabular}{lr}
        \toprule
        \textbf{Entry Type} & \textbf{Count} \\
        \midrule
        Code Intent & 2560 \\
        Complete the Code & 221 \\
        Inline Comments & 193 \\
        Header Comments & 130 \\
        Q\&A & 2877 \\
        \midrule
        \textbf{Total} & \textbf{5981} \\
        \bottomrule
    \end{tabular}
    \label{tab:dataset_task_type_breakdown}
\end{table}

\subsection{Model Selection}
\label{model-selection}

\begin{table}[ht]
    \centering
    \caption{LLMs selected for fine-tuning}
    \begin{tabular}{lcc}
        \toprule
        Model Series & Parameter Count & Quantization \\
        \midrule
        Qwen2.5-Coder & 3B & None \\
        Qwen2.5-Coder & 7B & None \\
        Qwen2.5-Coder & 14B & 4 bit \\
        Qwen2.5-Coder & 32B & 4 bit \\
        \midrule
        CodeLlama & 7B & None \\
        CodeLlama & 13B & 4 bit \\
        CodeLlama & 34B & 4 bit \\
        \midrule
        CodeGemma & 7B & None \\
        
        \bottomrule
    \end{tabular}
    \label{tab:selected_models}
\end{table}

As shown in Table \ref{tab:selected_models}, eight open-weight models spanning three LLM series are selected for fine-tuning. At the time of writing, these LLMs ranked near the top of HuggingFace's Big Code Model Leaderboard, making them suitable candidates for fine-tuning for code comprehension \cite{bigcode_leaderboard}. We select models between 3B and 34B parameters, a range which encompasses models with sufficient reasoning capability that also fit on consumer hardware. From the Qwen2.5-Coder series \cite{hui2024qwen2}, the 3B, 7B, 14B, and 32B models are selected. The 7B, 13B, and 34B models are used from the CodeLlama series \cite{roziere2023codellama}. Only the 7B model is selected from the CodeGemma series \cite{team2024codegemma} since there are no other CodeGemma models in the chosen parameter range that are compatible with the fine-tuning framework.

Note that in this paper, we focus on open-weight models trained for code understanding, that can run locally on an NVIDIA's RTX 5090, a high-end GPU intended for gaming. Our focus on local, open-weight models stems from concerns of privacy, security, and confidentiality. In many RE environments, internet access is completely disabled and/or strictly prohibited. 

Additionally, when dealing with malware, sending data over a network to a 3rd party API-based LLM is reckless, potentially risking accidental execution or exfiltration of malicious code. This same sentiment also applies to data privacy. Reverse engineered binaries could contain proprietary or sensitive information that must remain confidential, meaning such information should not be sent to 3rd party services that might collect this data. For these reasons, we argue that for most critical RE use cases, it is impractical to utilize external 3rd party API-based LLM systems, such as OpenAI's ChatGPT. REx86 can operate at full numerical precision on a consumer-grade PC equipped with an NVIDIA RTX 5090, with a token generation rate of over 100 tokens/second, requiring no internet access or data transmission out of the machine.

\subsection{Fine-tuning using Unsloth}


All fine-tuning described in this paper is conducted using the Unsloth framework. Developed in 2023, Unsloth is a framework that optimizes the fine-tuning process by reducing VRAM requirements and decreasing training time \cite{unsloth}. Their free, open-source framework employs a number of PEFT optimizations to facilitate two times faster fine-tuning with 50\% less VRAM usage \cite{unsloth}. Unsloth supports fine-tuning for a variety of popular LLM series such as Llama 4, Gemma 3, Phi 4, Mistral, and Qwen 2.5 \cite{unsloth}.

Some techniques used to expedite fine-tuning are flash attention, manual autograd optimization, custom OpenAI Triton kernels, LoRA, and quantization \cite{unsloth2024introducing}. Flash attention increases the efficiency of the attention mechanism in LLMs, which can often be a bottleneck because of the I/O required to transfer matrices between devices on a GPU \cite{dao2022flashattention}. Flash attention reduces the frequency that the key, value, and query matrices need to be transferred in the attention mechanism \cite{dao2022flashattention}. Unsloth also provides a custom implementation of PyTorch's autograd function, altering the order of matrix multiplication operations to optimize for LoRA fine-tuning. Additionally, Unsloth rewrites all kernels using OpenAI's Triton language  \cite{unsloth2024introducing}.

On top of this, Unsloth offers quantized models, allowing large LLMs to fit on consumer-grade hardware. It also utilizes LoRA to control the amount of trainable parameters and to allow easy distribution of fine-tuned models. 

Unsloth enables fine-tuning of larger models with less hardware, making it an ideal framework for our use case. It offers downloads for several state-of-the-art open-weight LLMs alongside documentation and code examples utilizing their models. Thus, we choose Unsloth as the fine-tuning framework for all experiments conducted in this paper.

\subsection{Process}

As mentioned in Section \ref{dataset_curation}, the fine-tuning dataset is split into training, validation, and testing sets. Each model is fine-tuned on a single NVIDIA RTX 6000 GPU with 48GB of VRAM using the hyperparmeters listed in Table \ref{tab:hyperparameters}. Each model undergoes fine-tuning on six different LoRA configurations in order to optimize LoRA hyperparameters. Specifically, the models are fine-tuned using a LoRA rank of 8, 16, and 32 with a 1:1 and 2:1 $\alpha:r$ ratio for each rank value. Each iteration consists of three epochs of training with validation run every ten steps. 

\begin{table}[]
    \centering
    \caption{Hyperparameters used for Fine-Tuning}
    \begin{tabular}{lr}
        \toprule
        Hyperparameter & Value \\
        \midrule
         Epochs & 3 \\
         Batch Size & 4 \\
         Gradient Accumulation Steps & 4 \\
         Weight Decay & 0.01 \\
         Learning Rate & 0.0001 \\
         Optimizer & adamw\_8bit \\
         Warmup Steps & 50 \\
         Evaluation Steps & 10 \\
        \bottomrule
    \end{tabular}
    \label{tab:hyperparameters}
\end{table}


\section{Evaluation}\label{sec:evaluation}




After fine-tuning, the models are evaluated quantitatively using cross-entropy loss and semantic cosine similarity on the testing dataset. The model with the highest holistic quantitative performance is then selected and referred to as REx86. REx86 is then evaluated through both a user study and also qualitatively.

\subsection{Fine-Tuning Results}

As discussed in Section~\ref{sec:methodology}, each model was trained for three epochs with a variety of LoRA hyperparameter configurations. Our fine-tuned model files are available at \url{https://zenodo.org/records/15420461}. Table \ref{tab:model_comparison} shows the optimal LoRA configuration and the minimum validation loss for each model. For every model, a $r$ of 32 and an $\alpha$ of 64 yielded the lowest validation in training. Figure \ref{fig:loss-graphs} in Appendix \ref{appendix} shows the validation and training loss for each model. Note that validation loss reaches a minimum at the conclusion of the second epoch for all models. After the second epoch, validation loss increases while training loss continues to decrease, which suggests the models begin overfitting beyond 2 epochs, which is typical for LLM fine-tuning. We select the model checkpoint at the minimum validation loss step for evaluation of each model.

With the exception of the CodeGemma model, the loss graphs for most models follow a similar pattern. CodeGemma's training and validation loss start much higher than the rest of the models, suggesting this model had very little prior exposure to x86 assembly. Despite the high initial loss, CodeGemma-7B converges to a loss value in the same range as the other models. 

\begin{table}[t!]
    \centering
    \caption{Fine-tuned model minimum validation loss comparison}
    \begin{tabular}{lccc}
        \toprule
        Model Name & Min Validation Loss & Rank  & Alpha \\
        \midrule
        Qwen2.5-Coder-3B & 0.68204 & 32 & 64 \\
        Qwen2.5-Coder-7B & 0.55780 & 32 & 64 \\
        Qwen2.5-Coder-14B & 0.62200 & 32 & 64 \\
        Qwen2.5-Coder-32B & 0.60006 & 32 & 64 \\
        \midrule
        CodeLlama-7B & 0.50468 & 32 & 64 \\
        CodeLlama-13B & 0.47266 & 32 & 64 \\
        CodeLlama-34B & 0.46085 & 32 & 64 \\
        \midrule
        CodeGemma-7B & 0.61131 & 32 & 64 \\
        \bottomrule
    \end{tabular}
    \label{tab:model_comparison}
\end{table}

\subsection{Quantitative Evaluation}
\label{mathematical_evaluation}

Through quantitative evaluation, we aim to answer \textbf{RQ1} by identifying the optimal LLM series and parameter sizes for generating comments for x86 code. We evaluated each model quantitatively using cross entropy loss (\texttt{CE}) and semantic embedding cosine similarity (\texttt{CosSim}), which are described further in this section. Table \ref{tab:base_vs_fine_tuned} shows a comparison between the base model and fine-tuned model metrics. Additionally, for more granular analysis, Table \ref{tab:breakdown_cos_sim} breaks down \texttt{CosSim} for each dataset task. In this section, we first outline the metrics used for quantitative analysis and then compare the strengths and weaknesses of the models in each LLM series.

\begin{table*}[h]
    \centering
    \caption{Comparison of base models to their fine-tuned counterparts on test set cross entropy loss and cosine similarity. $\Delta$\% indicates relative change from the base model.}
    \begin{tabular}{lcc|cc|cc}
        \toprule
        \textbf{Model Name} 
        & \multicolumn{2}{c|}{\textbf{Base Model}} 
        & \multicolumn{2}{c|}{\textbf{Fine-Tuned Model}} 
        & \multicolumn{2}{c}{\textbf{$\Delta$\% (FT vs Base)}} \\
        \cmidrule(lr){2-3} \cmidrule(lr){4-5} \cmidrule(lr){6-7}
        & \textbf{CE} & \textbf{CosSim} 
        & \textbf{CE} & \textbf{CosSim} 
        & \textbf{CE} & \textbf{CosSim} \\
        \midrule
        Qwen2.5-Coder-3B  & 1.94984 & 0.39552 & 0.66006 & 0.49081 & \textbf{-66.1\%} & \textbf{+24.1\%} \\
        Qwen2.5-Coder-7B  & 1.50482 & 0.55608 & 0.53901 & 0.66890 & \textbf{-64.2\%} & \textbf{+20.3\%} \\
        Qwen2.5-Coder-14B & 1.78891 & 0.56453 & 0.61010 & 0.46950 & \textbf{-65.9\%} & \textbf{-16.8\%} \\
        Qwen2.5-Coder-32B & 2.03961 & 0.49002 & 0.59197 & 0.55218 & \textbf{-71.0\%} & \textbf{+12.7\%} \\
        \midrule
        CodeLlama-7B      & 2.24458 & 0.51340 & 0.47708 & 0.69189 & \textbf{-78.7\%} & \textbf{+34.7\%} \\
        CodeLlama-13B     & 2.13460 & 0.43549 & 0.46332 & 0.45155 & \textbf{-78.3\%} & \textbf{+3.7\%} \\
        CodeLlama-34B     & 2.03961 & 0.45939 & 0.44231 & 0.54805 & \textbf{-78.3\%} & \textbf{+19.3\%} \\
        \midrule
        CodeGemma-7B      & 5.41661 & 0.16441 & 0.54475 & 0.17235 & \textbf{-89.9\%} & \textbf{+4.8\%} \\
        \bottomrule
    \end{tabular}
    \label{tab:base_vs_fine_tuned}
\end{table*}

\begin{table*}[h]
    \centering
    \caption{Comparison of cosine similarity across the five dataset subtypes for each fine-tuned model.}
    \begin{tabular}{lccccc}
        \toprule
        Model Name & Code Intent & Header & Inline & Completion & Q\&A\\
        \midrule
        Qwen2.5-Coder-3B & 0.47467 & 0.48625 & \textbf{0.52858} & 0.29873 & 0.51776 \\
        Qwen2.5-Coder-7B & 0.76260 & \textbf{0.59944} & 0.48088 & 0.53157 & \textbf{0.61230} \\
        Qwen2.5-Coder-14B & 0.42628 & 0.50408 & 0.48682 & 0.46367 & 0.50560\\
        Qwen2.5-Coder-32B & 0.53212 & 0.49190 & 0.49334 & 0.51714 & 0.57942 \\
        \midrule
        CodeLlama-7B & \textbf{0.82537} & 0.49779 & 0.46066 & 0.45744 & 0.48900 \\
        CodeLlama-13B & 0.40842 & 0.48392 & 0.31882 & 0.43480 & 0.49864 \\
        CodeLlama-34B & 0.51637 & 0.54307 & 0.49560 & \textbf{0.53958} & 0.58061\\
        \midrule
        CodeGemma-7B & 0.20980 & 0.33743 & 0.30231 & 0.25761 & 0.11626 \\
        \bottomrule
    \end{tabular}
    \label{tab:breakdown_cos_sim}
\end{table*}

\subsubsection{Quantitative Evaluation Metrics}
\label{quantitative_metrics}

The two metrics used to quantitatively evaluate the effectiveness of the fine-tuned LLMs are cross-entropy loss (\texttt{CE}) and semantic embedding cosine similarity (\texttt{CosSim}). \texttt{CE} compares the difference between the token probability distribution predicted by the model, $Q_i$, and the true token probability distribution for each token in a sequence, $P_i$. For each token in a sequence, $P_i$ is a one-hot encoded vector of length $V$, where $V$ is the model's vocabulary size and the correct token has probability of 1. 

For each sequence \( i \) in the test set, we compute the cross-entropy loss across all tokens, sum the losses, and normalize by the sequence length \( T_i \). We then average the normalized loss over all \( N \) sequences in the test set:

\[
\text{Cross-Entropy Loss}(P, Q) = - \frac{1}{N} \sum_{i=1}^{N} \left( \frac{1}{T_i} \sum_{j=1}^{T_i} P_{ij} \log Q_{ij} \right)
\]

Cross-entropy loss can be a useful general metric for evaluating how closely a model's responses align with a reference dataset. When comparing two models' testing dataset cross-entropy loss, the model with the lower loss has stronger alignment with the ground truth data. However, because it operates on a token-by-token basis, it penalizes the model even for semantically correct responses with different phrasing or word order. As a result, high loss values may not always reflect poor model performance. Therefore, additional metrics are needed to provide a more accurate and holistic assessment of model effectiveness.

Semantic embedding cosine similarity accounts for the semantic meaning of a sequence by comparing the semantic embedding vectors of the predicted and target sequences. These embeddings are high-dimensional vectors that encode the semantic content of a text segment, enabling meaningful comparisons in a multi-dimensional space. Cosine similarity, a common distance metric for comparing embeddings, measures the cosine of the angle between two vectors, $\mathbf{A}$ and $\mathbf{B}$, and is computed as: \[
\text{Cosine Similarity}(\mathbf{A}, \mathbf{B}) = \frac{\mathbf{A} \cdot \mathbf{B}}{\|\mathbf{A}\| \|\mathbf{B}\|}
\] Cosine similarity ranges from $-1$ to $1$, where $1$ indicates identical vectors, $0$ indicates orthogonal (unrelated) vectors, and $-1$ indicates completely opposing vectors.

In this work, we use NVIDIA’s open-weight NV-Embed-v2 model to generate embeddings for both model outputs and ground truth answers. At the time of writing, this model ranks among the top performers on the Hugging Face Embedding Leaderboard~\cite{mteb_leaderboard}.

The following sections analyze the strengths and weaknesses of the models of each LLM series based on (\texttt{CE}) and (\texttt{CosSim}) 

\subsubsection{Qwen2.5-Coder Series}
\label{qwen-quantitative-analysis}

Qwen2.5-Coder is the only series where a 3B parameter model is evaluated. This model series exhibits strong improvement after fine-tuning, especially with inline commenting, where it achieves the highest \texttt{CosSim} across all eight models. It struggles, however, with code completion tasks, yielding the second-lowest score in that category. 

The Qwen 7B model is perhaps the most balanced model after fine-tuning, yielding the highest \texttt{CosSim} for header commenting and Q\&A and the second highest \texttt{CosSim} for code intent samples. Its overall \texttt{CosSim} is exceeded only by the 7B CodeLlama model. 

The \texttt{CosSim} of the 14B Qwen model drops significantly after fine-tuning. Its \texttt{CE} value is also the second highest after fine-tuning. We observe that Qwen2.5-Coder-14B is relatively balanced across the dataset tasks with its worst task category being Code Intent. It is unclear why this specific model did not respond well to fine-tuning.

The Qwen2.5-Coder-32B model performs well in Q\&A but achieves average results in the remaining task categories. 

\subsubsection{CodeLlama Series} 
\label{codellama-quantitative-analysis}

CodeLlama-7B achieves the highest overall \texttt{CosSim} after fine-tuning, which is primarily attributed to its high \texttt{CosSim} for Code Intent tasks. The model's performance is average for all other tasks.

With the CodeLlama-13B model, we see a greater reduction in \texttt{CE} compared to the Qwen 14B model, which is consistent with the trend of other model sizes. However, CodeLlama-13B shows little improvement in \texttt{CosSim} after fine-tuning. A model's \texttt{CE} can improve without improving \texttt{CosSim} by learning to match the formatting of the ground truth data, but missing the semantic meaning. Note that this model performs poorly in inline commenting after fine-tuning and does not excel in any particular area. 

The 34B model is the largest model of the eight selected for fine-tuning. It improves more than its Qwen counterpart in terms of both \texttt{CE} and \texttt{CosSim}. It also achieves the highest code completion \texttt{CosSim}. 

\subsubsection{CodeGemma Series}
\label{codegemma-quantitative-analysis}

The CodeGemma model improves only marginally through the fine-tuning process. Its \texttt{CosSim} values do not exceed 0.35 for any task. While CodeGemma fails to excel in any task, it particularly struggles with Q\&A, harming its overall \texttt{CosSim}. Inspection of the model's outputs reveals that it fails to provide a response for the majority of Q\&A prompts with both the base and fine-tuned models. We attribute its poor performance on code-oriented tasks to a lack of x86 assembly exposure in pretraining. 

\subsubsection{Comparison}
\label{quantitative-analysis-comparison}

 The \texttt{CE} values for each of the eight models decrease significantly after fine-tuning. We observe that the CodeLlama models generally reach the lowest minimum \texttt{CE} in their respective parameter classes during fine-tuning. Additionally, Qwen2.5-Coder-7B and CodeLlama-7B achieve the highest \texttt{CosSim} over the test dataset after fine-tuning. For five of the eight models, fine-tuning results in a significant boost in \texttt{CosSim}. Fine-tuning CodeGemma-7B and CodeLlama-13B yield only a slight improvement and we actually see a decrease in \texttt{CosSim} after fine-tuning the 14B Qwen model. 

 Because of their high \texttt{CosSim} scores, Qwen2.5-Coder-7B and CodeLlama-7B are the highest performing models. While the CodeLlama model achieves the highest \texttt{CosSim} in code intent, it is outperformed by Qwen2.5-Coder in every other category. Thus, \textbf{we name the fine-tuned Qwen2.5-Coder-7B model REx86}, and it is used for subsequent human study evaluation and qualitative evaluation.
  

\subsection{User Case Study}
\label{human_evaluation}

We conducted a human case study to assess and compare the performance of REx86 against the base Qwen2.5-Coder-7B model in a reverse engineering (RE) context. This study was reviewed and approved by the Institutional Review Board (IRB) and determined to be exempt under Category 1 (minimal risk). 

\subsubsection{Study Design}

To emulate a realistic RE task, participants were presented with a fictional scenario in which the government has outlawed all references (textual or visual) to small mammals measuring less than one foot in length. Any mention of these creatures, regardless of spelling, was strictly forbidden. Participants were tasked with determining the intent of a newly discovered malware sample that is potentially related to this fictional legislation.
\subsubsection*{Crafted Malware Specimen (Windows/x86}\label{sec:case_study_specimen}

A custom malware sample was developed in C that performed various overt and covert actions to falsely implicate a computer user who executes the malware. Specifically, the malware:
\begin{itemize}
    \item Modified the Windows registry to insert fabricated squirrel-related URLs into Internet Explorer's history.
    \item Created a local directory and saved a ROT13-encoded manifesto expressing admiration for squirrels.
    \item Created a new user account named ``squirrellover'' and attempted to add it to the system administrator group.
    \item Wrote a squirrel image to disk and opened it multiple times using the default image viewer.
\end{itemize}

The specimen is a small Windows/x86 program written in C with two functions: \texttt{main} and a helper \texttt{k} that applies ROT13 to decode string constants at runtime. On execution it performs four actions: (1) seeds Internet Explorer’s \emph{TypedURLs} under \texttt{HKCU}, (2) creates a directory and writes a short text “manifesto,” (3) attempts local account creation and privilege escalation via shell commands, and (4) writes an embedded JPEG to disk and opens it repeatedly. There is no network activity, no autorun or persistence mechanism beyond the created artifacts, and all registry writes are under \texttt{HKCU} rather than \texttt{HKLM}.

The C source code and header files are available at: \href{hhttps://github.com/jghawaly/REx86-Squirrel-EvalSample}{https://github.com/jghawaly/REx86-Squirrel-EvalSample}.

\subsubsection*{Rationale for a crafted specimen}
We used a crafted sample to ensure a known, instruction- and function-level ground truth suitable for supervised comparison of header and inline comments, to allow safe distribution in a teaching lab, and to keep the task bounded to a three-hour session. The specimen was designed to be deterministic, to avoid network side effects, and to write only under \texttt{HKCU} and a controlled directory.




Participants were divided into three groups: REx86, Base, and Control. Each group received an IDA-64 file of the decompiled malware. The REx86 and Base groups were provided with versions of the IDA-64 file containing header and inline comments generated by their respective models (REx86 or Qwen2.5-Coder-7B). The Control group received no model-generated comments in their IDA-64 file. Additionally, the REx86 and Base groups were given access to a web-based chat interface to interact with their assigned model. Participants were assigned to groups randomly, with 75\% of participants being evenly split into the REx86 and Base model groups and the other 25\% being assigned to the control. This disproportionate assignment to the LLM groups was done in order to build up a higher sample size for the final survey, which was not applicable to the control group.

Before starting the task, participants completed a pre-survey and were briefed on the scenario. They were then given three hours to analyze the sample and determine its intent. Following the task, participants completed a post-survey evaluating the AI’s usefulness. Responses were collected using a 5-point Likert scale for the following items:
\begin{enumerate}
    \item The AI model’s output improved my understanding of the malware sample’s overall functionality. 
    \item The AI model’s output improved my understanding of the functionality of individual lines of Assembly code. 
    \item The responses of the chat model were helpful in the reverse engineering process. 
\end{enumerate}

\subsubsection{Participants}

Participants were upper-level undergraduate cybersecurity students enrolled in a malware reverse engineering course at a public R1 university in the United States. The course takes a hands-on approach, focusing on real malware samples and detailed commenting of disassembled code. The REx86 experiment was conducted in week 13 of a 16-week semester. Participation was voluntary, and data were collected only from students who provided informed consent. A total of 63 students consented to participate in the study. 

\subsubsection{Data Analysis}

Of the 63 submissions received, 3 were omitted due to empty surveys. An additional 17 were excluded based on poor performance in prior RE assignments, leaving 43 submissions for data analysis (18 in the Base group, 16 in REx86, and 9 in Control).

Mean Likert scores were calculated for each survey item. A one-tailed Mann-Whitney U test was used to evaluate whether the REx86 group reported significantly higher agreement than the Base group. Additionally, submissions were labeled as ``solved'' if the participant correctly identified the malware’s intent, i.e., to incriminate the user by accessing and generating squirrel-related content. Fisher's exact test was used to assess significance in solve rates across groups. The mean Likert scores and statistical comparisons for the base and REx86 groups are presented in Table~\ref{tab:likert_results}. 

\begin{table}[t!]
\centering
\caption{Likert means and solve rates by group. P-values correspond to one-tailed Mann--Whitney U (Likert) or Fisher's exact test (Solve). Statistically significant values ($p < 0.05$) in \textbf{bold}.}
\begin{tabular}{lccc}
\toprule
\textbf{Measure} & \textsc{Base} & \textsc{REx86} & \textsc{Control} \\
\midrule
\textbf{Sample Size (n)}             & 18         & 16         & 9         \\
\addlinespace
\textbf{Overall Understanding}       & $-0.67$    & $-0.13$    & --        \\
\quad $p$ (REx86 \ensuremath{>} Base)      & \multicolumn{3}{r}{$0.155$} \\
\addlinespace
\textbf{Line-Level Understanding}    & $-0.39$    & $0.50$     & --        \\
\quad $p$ (REx86 \ensuremath{>} Base)      & \multicolumn{3}{r}{\textbf{0.031}} \\
\addlinespace
\textbf{Chat Helpfulness}            & $-0.06$    & $0.06$     & --        \\
\quad $p$ (REx86 \ensuremath{>} Base)      & \multicolumn{3}{r}{$0.866$} \\
\addlinespace
\textbf{Solve Rate (\%)}             & 31.25      & 53.33      & 33.33     \\
\quad $p$ (REx86 \ensuremath{>} Base)      & \multicolumn{3}{r}{$0.189$} \\
\quad $p$ (REx86 \ensuremath{>} Control)   & \multicolumn{3}{r}{$0.916$} \\
\bottomrule
\end{tabular}
\label{tab:likert_results}
\end{table}

\subsubsection{Limitations} \label{sec:case_study_lim}
This human evaluation presents an empirical case study conducted in a teaching laboratory rather than a controlled professional user study. The cohort consisted of advanced students and was modest in size. To accommodate a fixed three-hour session, we evaluated a single malware specimen. Results may also be influenced by features of the lab environment, including familiarity with the tools and prompts. For these reasons, we do not claim broad generality; instead, we offer the findings as initial evidence of feasibility and utility and encourage replication with professional analysts using additional, real-world specimens.
\subsubsection{Discussion}

The human case study results indicate that REx86 offers improvements over the base Qwen2.5-Coder-7B model in certain aspects of the reverse engineering process. While both models were rated similarly in terms of overall intent analysis and chat assistance, the REx86 group reported significantly higher agreement when asked whether the model helped them understand the functionality of individual lines of Assembly code. This finding indicates that REx86 is particularly effective at providing localized, context-aware insights within disassembled code, which is a core task in the RE process. 

The REx86 group also showed a higher malware solve rate (53.33\%) than both the base model group (31.25\%) and the control group (33.33\%). While this difference did not reach the desired p-value, the trend aligns exactly with the expectation that REx86 should improve efficiency in the malware reverse engineering process. This result is very encouraging and studies with larger population sizes to better understand the significance are warranted, albeit very challenging to organize at scale in this particular area. This result suggests that the improved line-level support offered by REx86 over the base model may contribute to more actionable understanding, even if it does not always lead to a complete solution. Future studies with larger sample sizes will confirm this hypothesis to higher significance.

The lack of statistically significant differences in some areas may stem from the relatively small group sizes and the complex nature of reverse engineering tasks, where even minor gains in understanding can be impactful but difficult to quantify through simple metrics. Additionally, being the first time these participants interacted with these models for RE, more experience with using them and learning effective prompt engineering strategies may yield different results.

Overall, despite the limitations in this human evaluation study, these results signalize REx86’s value as a tool that augments human reverse engineers by improving clarity and comprehension at the micro (line-by-line) level. While broader gains in overall performance and solve rate remain an area for future exploration through larger studies, the data suggests support for REx86’s role in enhancing RE productivity and understanding through more interpretable and actionable model outputs.

\subsection{Qualitative Evaluation}
\label{qualitative_analysis}


In this section, we examine two snippets of x86 assembly code from the test dataset and compare the performance of REx86 to the base Qwen2.5-Coder-7B model. We select a bitwise operation snippet, and an obfuscated snippet to evaluate. This section aims to answer \textbf{RQ2} by evaluating the effectiveness of local, open-weight LLMs in understanding x86 assembly code.


\subsubsection{Bitwise Operation Sample}
\label{bitwise_operations_sample}

The first snippet sample is a snippet of x86 code that uses various bitwise operations to swap different amounts of bits in the \texttt{EAX} register. Code \ref{bitwise_code_snippet} shows the full code sample provided provided to both the base model and REx86.

\begin{codebox}[Bitwise Operations Example]
\label{bitwise_code_snippet}
\begin{lstlisting}[
    frame=none, 
    aboveskip=0pt,
]
SECTION .text
	global main
main:
	mov	ecx,eax		
	and	ecx,0xffff0000	
	shr	ecx,0x10	
				
	and	eax,0x0000ffff	
	shl	eax,0x10	
				
	or	eax,ecx		
	
	mov	ecx,eax		
	and	ecx,0xff00ff00	
	shr	ecx,0x8		
	and	eax,0x00ff00ff	
	shl	eax,0x8		
	or	eax,ecx		
	
	mov	ecx,eax		
	and	ecx,0xcccccccc	
	shr	ecx,0x2		
	and	eax,0x33333333	
	shl	eax,0x2		
	or	eax,ecx		
	
	mov	ecx,eax		
	and	ecx,0xf0f0f0f0	
	shr	ecx,0x4		
	and	eax,0x0f0f0f0f	
	shl	eax,0x4		
	or	eax,ecx		
	
	mov	ecx,eax		
	and	ecx,0xaaaaaaaa	
	shr	ecx,0x1		
	and	eax,0x55555555	
	shl	eax,0x1		
	or	eax,ecx	
\end{lstlisting}
\end{codebox}

\begin{commentbox} {Bitwise Operation Snippet Comments}\label{comments:bitwise}

\textbf{Qwen2.5-Coder-7B}

The assembly snippet is designed to perform a series of bitwise operations on the input value stored in the EAX register. The goal appears to be manipulating the bits of the input value in a specific pattern, likely for encryption or data transformation purposes. Here's a header comment summarizing its functionality \\

\textbf{REx86}

This snippet takes the contents of the eax register, extracts the high and low 16 bits, and reverses their order in the result. It does this using a sequence of and, shl, shr, and or instructions. The same is done three more times, each time reversing the order of the high and low 8 bits, the high and low 4 bits, and finally the high and low 2 bits. The final result is stored in eax. For eax = 1, it returns 1717986919.
\end{commentbox}

After evaluating several responses from the base LLM and REx86 on the bitwise operation sample, we see a clear improvement in response quality with REx86, as shown in Comment \ref{comments:bitwise}. Specifically, the base Qwen model generates very vague descriptions of the code's functionality, pointing out that the instructions that the snippet contains and asserting that the program ``manipulate its bits and perform a specific transformation''. The base model also points to encryption or decryption as the intent of the snippet, which leads the user astray and fails to encapsulate the point of the instructions given. 

REx86, on the other hand, is consistently able to identify that this snippet swaps different amounts of bits within the EAX register. It uses more precise language and refrains from offering conjectures that distract from the intent of the snippet. Note that REx86 still confuses the order of the swaps in its description. It is mistaken in stating that the 4-bit swap occurs before the 2-bit swap. However, its description still facilitates effective understanding of the code rather than a vague description of instructions used.



\subsubsection{Obfuscated Code Sample}
\label{obfuscated_code_sample}

The second sample tested demonstrates a variety of unnecessarily complicated ways to sway the values of the EAX and ECX registers as shown in Code \ref{obfuscated_code_snippet}.

\begin{codebox}[Obfuscated Register Swap Snippet]
\label{obfuscated_code_snippet}
\begin{lstlisting}[
    frame=none, 
    aboveskip=0pt,
]    
SECTION .text
	global main
main:
	push	rax		
	push	rcx		
	pop	rax		
	pop	rcx		
	xor	rax,rcx		
	xor	rcx,rax		
	xor	rax,rcx		
	add	rax,rcx		
	sub	rcx,rax		
	add	rax,rcx		
	neg	rcx		
	xchg	rax,rcx	
\end{lstlisting}
\end{codebox}

\begin{commentbox} {Obfuscated Sample Comments}\label{comments:obfuscated}

\textbf{Qwen2.5-Coder-7B}

Header Comment: The provided assembly snippet implements a series of bitwise XOR and arithmetic operations to perform a specific task. It manipulates the values of the RAX and RCX registers through a sequence of instructions that ultimately results in a modified value stored in RAX, while RCX is modified accordingly.\\

\textbf{REx86}

This snippet demonstrates the following interesting properties:\\
    1) xor rax,rcx; xor rcx,rax; xor rax,rcx; is equivalent to xchg rax,rcx; \\
    2) add rax,rcx; sub rcx,rax; add rax,rcx; is equivalent to neg rcx; xchg rax,rcx;

\end{commentbox}

This sample also heavily uses bitwise operations, and the base model defaults to similar vague language, suggesting limited comprehension of functionality. REx86 is able to identify that the purpose of the code is swapping register values in obscure ways. However, it fails to recognize the first method of swapping that utilizes the stack. Nevertheless, REx86 offers superior assistance in understanding this snippet compared to the base model. 

Overall, the general trends seen in evaluating REx86 are increased specificity, reduced hallucination, and a stronger grasp of the bigger picture.

\section{Limitations and Future Work}\label{sec:limits_future}

One limitation of this work is the size of the fine-tuning dataset, which is attributed to a lack of available data. While online repositories contain an abundance of x86 assembly code, there is a lack of commented disassembly output. Since disassembly output is central to most RE tasks, future work could explore developing a dataset of commented disassembly to improve the model's performance on full malware samples and firmware images. 

Additionally, this paper focuses solely on the x86 architecture. We leave it as future work to extend the model's functionality to other instruction sets, such as ARM and MIPS. We theorize that architecture generalization could be achieved through multiple avenues. One potential approach is to train an additional LoRA adapter for each new supported architecture. While this could provide the highest degree of specialization in each architecture, it would require a heavy initial lift for each new architecture. Additionally, a larger dataset with entries distributed over a variety of architectures could help the model comprehend different instruction sets. This generalized model could then retrieve specific syntactical information for different architectures from RAG databases. 

As detailed in Sec.~\ref{sec:case_study_lim}, our human evaluation was a teaching-lab case study with trained students (n=43) and a single specimen. Accordingly, the findings should be interpreted as indicative rather than generalized. Future work will pursue professional replications in operational environments, expand to multiple real-world specimens with ground-truth annotations, and evaluate longer-duration workflows.


\section{Ethical Considerations}\label{sec:ethics}



In this paper, we create REx86, an LLM specialized in assisting in RE. Since it is a generative model fine-tuned on x86 assembly, REx86 has a greater capacity for generating x86 code, even malicious code. Liguori et al. demonstrate the ability of a model to generate shellcodes, small snippets of code used to exploit a vulnerability in a piece of software \cite{ShellcodeIA32}. While they don't use LLMs, they train their model on a dataset similar to the one constructed in this paper. 

Additionally, REx86 assists in RE, lowering the barrier of entry for both well-intentioned practitioners and malicious actors. This stems from the fact that RE, like any other tool, can be used or abused in a variety of ways. For example, RE is used to discover and patch vulnerabilities in critical infrastructure, but it is also used by malicious actors to identify and subvert security measures or pirate software. We maintain that the positive use cases for REx86 outweigh any potential abuses of the model. 

\section{Conclusion}\label{sec:conclusion}

Reverse engineering is an arduous and time-consuming process due to several factors such as compiler optimizations and code obfuscation. LLMs possess a natural language prowess that could lend itself well to demystifying the reverse engineering process through contextualized comments and code summarization. However, LLMs often struggle with low-level code comprehension, so additional tuning is needed to maximize the effectiveness of LLMs in RE. 

In this paper, we fine-tune a variety of LLMs on a dataset of x86 assembly code to evaluate their effectiveness in assisting with reverse engineering. We focus on the popular x86 instruction set and select open-weight models suitable for running on consumer-grade hardware. To fine-tune the models, we use a framework called Unsloth that reduces the time and resource requirements for fine-tuning through a number of optimizations, such as LoRA and quantization. 

Results from quantitative evaluation, a human user case study, and qualitative analysis all indicate promising improvements. Five of the eight models showed significantly higher semantic embedding cosine similarity scores on the test set after fine-tuning. The best-performing model, REx86, is a fine-tuned version of Qwen2.5-Coder-7B. REx86 helped participants in the user study better understand the intent behind lines of disassembled code within a malware sample. Finally, qualitative analysis of selected examples highlights meaningful improvements in the output quality of REx86 over the base Qwen2.5-Coder-7B model.

Results demonstrate that REx86, while not capable of fully automating RE, provides superior assistance in reverse engineering efforts compared to current local, open-weight models by fostering a deeper understanding of x86 assembly code.


\bibliographystyle{IEEEtran}
\bibliography{references}

\appendix
\label{appendix}

\begin{figure*}
    \centering
    \includegraphics[width=\textwidth]{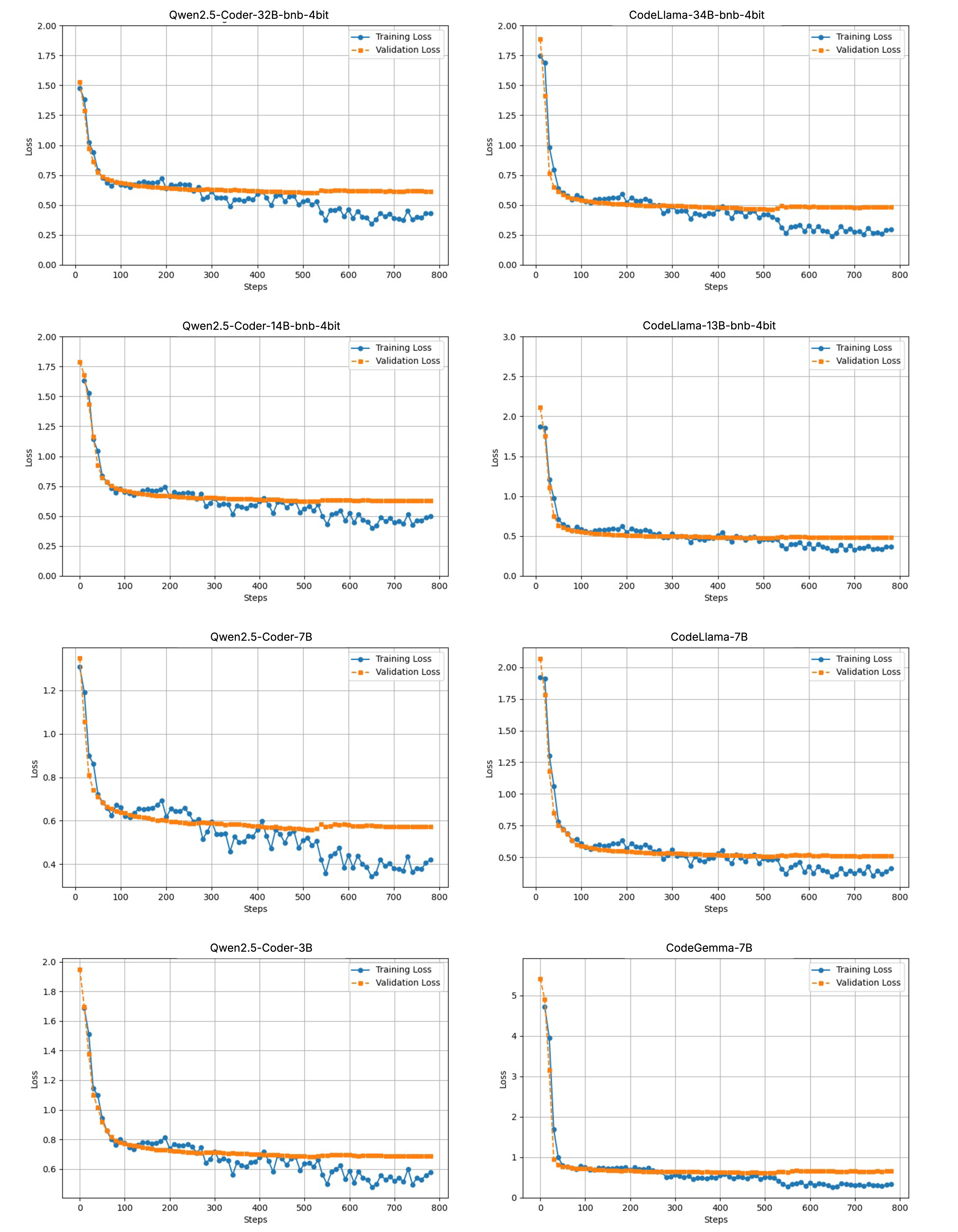}
    \caption{Training and validation loss for each of the 8 models throughout the training process.}
    \label{fig:loss-graphs}
\end{figure*}

Figure \ref{fig:loss-graphs} displays the loss graphs for the fine-tuning of each of the eight models using the optimal LoRA parameters outlined in \ref{tab:model_comparison}. The validation loss is plotted in orange alongside the training loss in blue. The figure allows us to identify the point at which the validation loss reaches a minimum and overfitting starts to occur. Note that the minimum validation loss is reached at the end of the second epoch for all eight models. 

\end{document}